# TA-Designed vs. Research-Oriented Problem Solutions

Shih-Yin Lin[1], Chandralekha Singh[1], William Mamudi[2], Charles Henderson[2], Edit Yerushalmi[3]

[1]*Department of Physics and Astronomy, University of Pittsburgh, Pittsburgh, PA 15213, USA*
[2]*Department of Physics and Mallinson Institute for Science Education, Western Michigan University, Kalamazoo, MI 49008, USA*
[3]*Department of Science Teaching, Weizmann Institute of Science, Rehovot, 76100, Israel*

**Abstract.** In order to study graduate teaching assistants (TAs) beliefs and values about the design of instructor problem solutions, twenty-four TAs were provided with different solutions and asked to discuss their preferences for prominent solution features. TAs preferences for solution features were examined in light of the modeling of expert-like problem solving process as recommended in the literature. Results suggest that while many of the features TAs valued align with expert-like problem solving approaches, they noticed primarily "surface features" of solutions. Moreover, self-reported preferences did not match well with the solutions TAs wrote on their own.

**Keywords:** problem solving, problem solutions, teaching assistants, teachers beliefs.

**PACS:** 01.40.Fk, 01.40.gb

## INTRODUCTION

Cognitive apprenticeship approach [1] underlies many pedagogical techniques that have been shown to promote expert-like problem solving. In this approach a prescribed problem-solving framework is made explicit through "modeling" it in instructors' solutions to problems. The framework involves: 1) initial problem analysis, 2) solution construction (choice of sub-problems), and 3) checking of solution [2].

If we wish to help instructors make problem solving approaches explicit on problem solutions they provide students, it is necessary to understand how these instructors currently perceive and value the design features of solutions to problems.

In previous work we have investigated faculty beliefs and values related to the use of instructor solutions [3,4]. In this paper, we report on an investigation of the beliefs and values of graduate teaching assistants (TAs). TAs play a central role in the teaching of physics problem solving in many physics departments. Two main research questions are:
(1) Do TAs notice and value features that explicate the expert decision-making process?
(2) What do TAs have in mind when "discussing/mentioning" features that explicate the expert decision-making process?

## METHODOLOGY

Twenty four first-year graduate TAs enrolled in a TA training course were provided with three instructor solutions for the same physics problem and asked to explain how these solutions compare with their preferences for the design of instructor solutions. Data were collected using a Group-Administered Interactive-Questionnaire (GAIQ) approach [5] in which each TA first wrote a solution for the designated problem that they would hand out to their students. The TAs then read three example problem solutions and identified prominent features of those solutions (e.g., providing a diagram) in a worksheet. They also ranked the three solutions based on a) which solution has more of each feature, and b) their preference for including these features in solutions. TAs were also asked to explain the reasons behind their preferences. To verify meaning and allow for the sharing of ideas, TAs were later asked to discuss their ideas in small groups and report their conclusions in a whole class discussion. Finally, each TA was given the opportunity to explain whether (and why) their preference changed by filling in a similar post-discussion worksheet. On this post-discussion worksheet they were asked to match the features they identified on the pre-discussion worksheet to a list of pre-defined features (See Table 1) representing different aspects of the solution presentation. The list represents categories of features identified in a pilot study with the same population. Some of these categories relate to the expert problem solving process [2]. Both the pre- and post-discussion worksheets as well as TAs' own solutions were collected for analysis. Features on the pre-worksheet that were not matched to Table 1 by the TAs were categorized as additional features by the researchers. The complete corpus of data was analyzed by two researchers. Any disagreements were discussed by 4 researchers until full agreement was established. The details of the GAIQ approach are presented in a companion paper [5].

**TABLE 1.** Pre-defined feature list (from pilot study).

| |
|---|
| 1. Provides a schematic visualization of the problem (a diagram) |
| 2. Provides a list of knowns/unknowns |
| 3. Provides a "separate" overview of how the problem will be tackled (Explains premise and concepts -- big picture -- prior to presenting solution details) |
| 4. Explicit sub-problems are identified (Explicitly identifies intermediate variables and procedures to solve for them) |
| 5. Reasoning is explained in explicit words (Description/justification of why principles and/or subproblems are appropriate/useful in this situation) |
| 6. The principles/concepts used are explicitly written using words and/or basic mathematical representations (e.g., F=ma or Newton's 2$^{nd}$ Law) |
| 7. Thorough derivation (Detailed/verbose vs. Concise/short/simplified/skips lots of derivation) |
| 8. Long physical length (Long/verbose vs. Short/concise vs. Balanced/not too long, not too short) |
| 9. Includes details that are not necessary for explaining the problem solution (The solution is technically correct and complete without these 'unnecessary' details) |
| 10. Provides alternative approach |
| 11. Solution is presented in an organized and clear manner |
| 12. Direction for the progress of the solution progress: Backward vs. forward |
| 13. Symbolic solution (Numbers are plugged-in only at the end) |
| 14. Provides a check of the final result (e.g. if the unit is correct, or if the answer makes sense by examining the limits) |

# FINDINGS

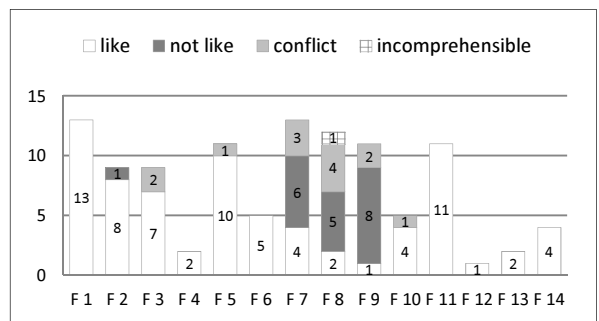

**FIGURE 1.** Number of TAs mentioning each feature.

In addition to the 14 pre-defined features given in Table 1, there were 3 additional features that the TAs noticed. Because each was mentioned by only 1 or 2 TAs, we will focus only on the pre-defined features. Figure 1 shows the number of TAs who noticed each of the pre-defined features, and whether or not they liked it or were conflicted about it. If the TAs' preference for the feature changed after the discussion, or if the TAs explained both the pros and cons of a feature, they are placed in the "conflict" category. In the following we will separate our discussion of these results as related to the different components of an expert-like problem solving process [2].

*1. Features Related to Initial Problem Analysis*

Providing a schematic visualization of the problem (F1) and providing a list of knowns/unknowns (F2) are the features that relate to the explication of the initial problem analysis stage in an expert-like problem solving process [2]. F1 is one of the most mentioned features (13 out of 24 TAs). F2 was mentioned by 9 TAs (the median for all features). These features were valued by almost all TAs who mentioned them. Only one TA expressed that he didn't like to provide a list of knowns/unknowns because it encourages students to solve problem via mindless plug and chug. Other TAs valued the list of knowns/unknowns because it "gives an idea of what you have and what you need." Examination of TAs' own solutions (which 23 TAs provided) indicates that all TA solutions included a diagram. The list of knowns (and sometimes with the unknown targeted variable included) was found in the solutions of 12 TAs.

Although all TAs valued F1 (visualization), different TAs had different ideas about the preferred visualization shown in Figure 2. Table 2 shows that initially 9/13 TAs distinguished between the quality of diagrams, with 6 of them preferring a detailed drawing as presented in solution 3. Most of the TAs did not articulate why the detailed diagram was better than the others. TAs who chose the less detailed diagrams in solution 1 and/or 2 explained, for example, that they didn't like diagram 3 because "complicated diagrams can be confusing".

Some TAs worried that the arrows in diagram 3 could be confusing to the students because they are used to represent both acceleration and velocity. It is likely that this concern was spread during the peer discussion stage, and therefore between the pre and the post the number of TAs who did not distinguish between solutions decreased and the number of TAs preferring solution 1 increased.

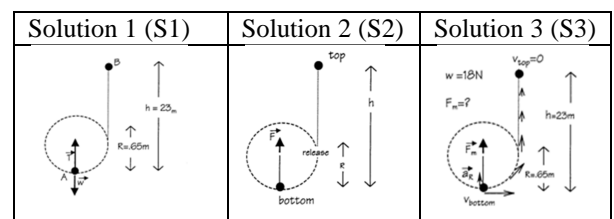

| Solution 1 (S1) | Solution 2 (S2) | Solution 3 (S3) |
|---|---|---|

**FIGURE 2.** Diagram used in each of the 3 solutions.

**TABLE 2.** TA's preferences for each type of diagram.

| Solution | Number of TAs (pre) | Number of TAs (post) |
|---|---|---|
| S1 | 1 | 3 |
| S2 | 1 | 1 |
| S3 | 6 | 5 |
| S1=S2 | 1 | 1 |
| S2=S3 | 0 | 1 |
| S1=S2=S3 | 4 | 2 |

## 2. Features Related to Solution Construction

Six of the features (F3, F4, F5, F6, F10, F12) relate to the solution construction stage in an expert-like problem solving process. They can be further classified into 3 groups shown below:
Choices made (major solution steps):
  F4) Explicit sub-problems are identified
  F6) Principles/concepts used are explicitly written
Reasons for choices (additional explanations):
  F3) Providing a "separate" overview
  F5) Reasoning is explained in explicit words
Framework within which choices are made:
  F10) Providing alternative approach
  F12) Forward vs. backward solution

Based on Reif's [2] suggestion to represent the process of solving a problem as a decision making process, the major choices a person makes in a solution process involve defining sub-problems: intermediate variables and principles to find them. Underlying these choices is the solver's reasoning. While F4 and F6 present the major choices one makes, F3 and F5 provide additional explanations regarding the reasons underlying these choices. We note that this reasoning is guided by the solver's general perception of the framework within which choices are made (e.g., as a process that involves choosing between alternatives, or arriving at identified goal in a backward manner) represented in F10 and F12. Figure 1 shows that features related to *reasons for choices* were the most noticed ones.

Table 3 shows the solutions TAs believed best represent features related to *reasons for choices*. Most of the TAs who noticed these features thought that they were best represented in solution 2 or 3. However, as shown in Figures 3 and 4, these solutions present reasoning in different ways. Solution 2 identifies the goal of each sub-problem and provides justification for the principles separately as the progress of the solution. Solution 3 describes a complete overview of how the problem should be broken into sub-problems and explains the principles applicable in each of the sub-problems at the very beginning. In general, solution 3 was slightly preferred by TAs for its enactment of F3 while solution 2 was generally preferred as the best enactment of F5. Although most TAs did not explicate why one presentation is better than the other in the worksheets, in the whole-class discussion several TAs raised their concerns that students may not

TABLE 3. TAs' preferences for F3 and F5.

|         | F3 (pre) | F3 (post) | F5 (pre) | F5 (post) |
|---------|----------|-----------|----------|-----------|
| S1      | 0        | 1         | 0        | 0         |
| S2      | 2        | 3         | 7        | 6         |
| S3      | 5        | 4         | 2        | 2         |
| S1=S3   | 1        | 1         | 0        | 0         |
| S2=S3   | 0        | 0         | 2        | 2         |
| S1=S2=S3| 1        | 0         | 0        | 1         |

Step 1) Find $v_r$ needed to reach h
$E_i = E_f$
$E_{release} = E_{top}$
$PE_{release} + KE_{release} = PE_{top} + KE_{top}$
$mgR + mv_r^2/2 = mgh + mv_t^2/2$
$v_r^2 = 2g(h - R)$

Conservation of energy for the stone earth system, since no external forces.
Note: you could also choose other systems.
KE of earth estimated to be 0
You could also use kinematics to find $v_r$.

**FIGURE 3.** Example presentation of F3 and F5 in S2.

Approach:
I need to find $F_m$, force exerted by me. I know the path, h (height at top) and $v_t$ (velocity at top)
A) For a massless string $F_m = T_b$ ($T_b$-Tension at bottom)
B) I can relate $T_b$ to $v_b$ (velocity at bottom) using the radial component of $\sum \vec{F} = m\vec{a}$, and radial acceleration $a_R = v^2/R$, since stone is in circular path
C) I can relate $v_b$ to $v_t$ using either i) energy ii) Dynamics and kinematics
    ii) Messy since forces/accelerations change through the circular path
    i) I can apply work-energy theorem for stone. Path has 2 parts:
        first - circular, earth and rope interact with stone,
        second - vertical, earth interacts with stone
    In both parts the only force that does work is weight, since in first part hand is not moving $\Rightarrow \vec{T} \perp \vec{v} \Rightarrow \vec{T}$ does no work.

**FIGURE 4.** Presentation of F3 and F5 in S3.

have the patience to read the whole chunk of text at the beginning of solution 3. Students may simply ignore all the explanations in the first part and jump directly into the second part with equations. Reasoning that is presented beside the equations, as in solution 2, makes it easier to reference and students are more likely to process the information better.

In general, F3 and F5 were valued by most TAs who noticed them. The TAs believed that these features play an important role in instructor solutions because they make the solution process clear and make the solution easier to follow. The TAs also believed that these features help students understand the internal thinking process that the instructor went through when solving the problem and facilitate better transfer to other problems. Except for minor concerns, such as "overdoing the motivations can lead to undesired chunks of text", which was the major reason why a few of the TAs expressed a conflicted preference, these features were generally valued by TAs. However, examination of TAs' own solutions indicates a discrepancy between their self-reported preferences and their actual practice. In total, only 3 out of 23 TAs provided some outline of the sub-problems (F3) either at the very beginning or along the solution progression, and only 6 of the 23 TAs provided any justification for the principles used (F5).

Features 4 and 6, which explicate the *choices made*, were less noticed (2 and 5 TAs, respectively), although they were valued by all TAs who noticed them. One TA explained that "I enjoy this feature [F4] because it helps set up a logical progression of the problem"; other TAs explained their preference towards F6 in that "the concepts may be more important than the answer" or "if we can use less math, I think we should do that, so students focus on physics". Examination of TAs' own solutions indicates that no TA presented a solution in which the goals for each sub-problem were clearly stated. On the other hand, the concepts of "conservation of energy" and "Newton's $2^{nd}$ Law" were explicitly written in words or the basic mathematical forms by 18 and 8 TAs, respectively.

Regarding the *framework within which choices are made*, 4 of the 5 TAs who noticed F10 (providing alternative approach) preferred this feature, explaining, for example, that "this [feature] demonstrates how to develop an expert knowledge structure and how it makes the problem much simpler." One TA was conflicted about this feature, as presenting an alternative approach "could possibly confuse students." However, no TA provided an alternative approach in their own solutions. As for F12 (backward vs. forward solution), most TAs did not notice it as an important consideration in the design of a solution. One difference between experts and novices is experts (teachers) commonly regard introductory physics problems as exercises while they are actually problems for novices (students). As a result experts may present problem solutions in a forward manner, reflecting their knowledge of the problem solution in an algorithmic way. Yet, to explicate the decision making process of an expert when solving a real problem, as suggested by instructional strategies aligned with cognitive apprenticeship [1], one has to present the solution in a backward manner. Only one TA mentioned this feature. However, this TA presented his/her solution in a forward manner. On the other hand, there were 8 TAs who originally presented a backward solution, even though they did not mention F12 in the worksheets. It is likely that many of the TAs consider the backward and forward solutions as interchangeable.

*3. Features Related to Checking of Solution*

F14, providing a check of the final result, is the feature which is related to the last step of an expert problem solving process: checking of solution. We expected this feature to stand out in the artifact comparison technique since only 1 of the 3 solutions included it. However, only 4 TAs noticed this feature. In addition, examination of TAs' solutions indicates that none of the TAs performed an answer check in the solutions they prepared for the introductory students. Although this feature was valued by all the TAs who noticed it, the findings suggest that this feature was underrated or ignored by most of the TAs.

## CONCLUSIONS

In general, we find that the TAs did notice and value features related to the explication of an expert-like problem solving process, in particular, problem re-description and the planning of the solution. Yet, most features that the TAs noticed were "surface features" such as F1 (drawing), F3 (separate overview), and F9 (length) that one is likely to be aware of even if s/he doesn't know much about physics problem solving. This is compared to features such as F6 (principles used) or F12 (direction) that are deeper features of the solution and were less commonly identified by the TAs.

In addition, we find that the self-reported preferences didn't match well with the solutions TAs wrote on their own before seeing the 3 artifacts. Although features in all 3 groups that are aligned with the expert-like problem solving process were in general valued by the TAs, only features related to problem re-description (especially F1) were generally found in their own solutions. The majority of the TA solutions contained little or no reasoning to explicate the underlying thought processes. No answer check was found in any TA's solution. We note that the TAs' solutions were collected at the beginning of the TA training course, when the TAs had just entered graduate school and started their TA jobs. It is likely that this activity, which helps to elicit TAs' initial ideas about the design of problem solutions in physics teaching, will influence their practices in the future.

Thus, we believe that the activity described in this paper provides a starting point for TAs' professional development. In addition to this activity, follow up activities that are aligned with the theoretical strategies for enhancing conceptual change could be implemented. For example, it would be beneficial if new ideas are imported from the research literature, and the TAs are explicitly guided to evaluate their practice in light of these new ideas.